# Rotational Bands and Electromagnetic Transitions of some even-even Neodymium Nuclei in J-Projected Hartree-Fock Model

S. K. Ghorui[1,*], Z. Naik[2], S. K. Patra[3], A. K. Singh[1], P. K. Raina[1,4], P. K. Rath[5] and C. R. Praharaj[3]

[1]*Department of Physics & Meteorology, IIT Kharagpur, Kharagpur-721302, INDIA*
[2]*Department of Physics, Sambalpur University, Burla-768019, INDIA*
[3]*Institute of Physics, Bhubaneswar-751005, INDIA*
[4]*Department of Physics, IIT Ropar, Rupnagar-140001, INDIA and*
[5]*Department of Physics, University of Lucknow, Lucknow-226007, INDIA*

(Dated: June 16, 2011)

Rotational structures of even-even $^{148-160}$Nd nuclei are studied with the self-consistent deformed Hartree-Fock (HF) and angular momentum (J) projection model. Spectra of ground band, recently observed $K = 4^-$, $K = 5^-$ and a few more excited, positive and negative parity bands have been studied upto high spin values. Apart from these detailed electromagnetic properties (like E2, M1 matrix elements) of all the bands have been obtained. There is substantial agreement between our model calculations and available experimental data. Predictions are made about the band structures and electromagnetic properties of these nuclei. Some 4-qasiparticle K-isomeric bands and their electromagnetic properties are predicted.

PACS numbers: 21.10.-k, 21.10.Ky, 23.20.lv, 27.70.+q

## I. INTRODUCTION

The study of rotational structures of neutron-rich nuclei is one of the main topics of nuclear physics. For rotational nuclei the coupling of single-particle and collective rotational degrees of freedom leads to a variety of phenomena [1–3]. The neutron number N=90 marks the beginning of deformed rotational features in the rare-earth region. This is seen in a number of nuclei in this region [4–6]. Lighter Neodymium isotopes show vibrational like behavior, while heavier isotopes display more rotational-like spectra [7]. In heavier Nd isotopes the active protons are filling in sdg$_{7/2}$h$_{11/2}$ shell above $Z = 50$ shell closure and active neutrons in pfh$_{9/2}$i$_{13/2}$ shell above $N = 82$ shell closure. Hence, these Nd nuclei are expected to show deformation properties and associated band structures. Some experimental informations [8–13] are now available for the band structures of neutron-rich Nd nuclei and more results can be expected with the existing [14–16] as well as new generation facilities [17–21] for structure studies away from the valley of stability towards the neutron drip line. It is thus of great interest to make microscopic study of the structure of these nuclei and to know quantitatively the bands to be expected for these.

For a reliable description of the band structure, one needs a microscopic many-body self consistent procedure to give insight into the shape and intrinsic structure of the nucleus and angular momentum projection procedure to obtain the various J states from the intrinsic states. The model used should be based on residual interaction among nucleons in a reasonable model space. In this work we use deformed Hartree-Fock and Angular Momentum (J) projection method [22–24] for the study of band structures of even-even Nd nuclei ($A = 148 - 160$). We call the deformed HF and angular momentum projection the Projected Hartree-Fock (PHF) model. This self-consistent procedure ensures the correct particle number (for protons and neutrons) and gives eigenstates of angular momenta for the states of the bands. This model is known to take into account the configuration mixing in the active shell model space and hence gives a quantitative description of energy spectra and electromagnetic transitions in deformed nuclei [25–27]. The residual interaction is included self-consistently in building the deformed basis in this theory. Once the self-consistent HF calculation has been performed, one can build the intrinsic states of the bands by particle-hole excitations across the proton and neutron Fermi surfaces (the various quasiparticle configurations), besides the Hatree-Fock configuration. Subsequent J-projection from the deformed intrinsic configurations gives the spectra of various bands. Diagonalisation after projection can be done. Thus our model is very close to the shell model [28] and can predict new bands with higher energies and higher angular momenta.

The neutron rich Nd nuclei are well deformed (rotational states are known in $^{148}$Nd-$^{156}$Nd). But the experimental information are quite limited. Only a few bands are known and they are not known to high spin values. No bands are know experimentally for the heavier neutron-rich Nd nuclei ($^{158}$Nd onwards). However, the current experimental facilities like spontaneous fission of $^{252}$Cf, thermal neutron induced fission make it possible to study band structures of heavier deformed rare-earth nuclei. The experiment with spontaneous fission of $^{252}$Cf is used by Zhang et al. [8] and

---

* email: surja@phy.iitkgp.ernet.in

Gautherin et al. [10] to study yrast band, negative parity bands and bands based on K-isomers in $^{152,154,156}$Nd nuclei. Recently, using the same technique, Simpson et al. [9] reported the two-quasiparticle (2-qp) isomeric band structure for $^{154,156}$Nd nuclei. They are based on $4^-$ isomeric state in $^{154}$Nd and $5^-$ isomeric state in $^{156}$Nd. Similar band structure based on $7^-$ isomeric state is also observed by Yeoh et al. for $^{152}$Nd, in their recent study [29]. Motivated by such a good number of experimental observations and very few systematic theoretical descriptions so far of these nuclei we have study the band structures for a wider range of even-even Nd nuclei isotopic chain with mass numbers $A = 148$ to $A = 160$ in a self-consistent model (the PHF model).

This paper is organized as follows: some details of the theoretical formalism and calculation procedure are presented in Section II. The results and discussion are given in Section III. The various possible intrinsic configurations used in our calculations are presented here. Comparisons of theoretical results with available experimental data for energy levels, bandhead energies and electromagnetic properties for even-even $^{148-160}$Nd are made. Finally we summarize our work in Section IV.

## II. DEFORMED HARTREE-FOCK AND J PROJECTION METHOD

### A. Deformed HF Formalism

The HF equation is derived from the nuclear Hamiltonian which consists of single-particle and residual two body interactions terms. The self-consistent HF method is a very useful technique to obtain nuclear mean field from two body interaction. Here we have used Surface Delta interaction (SDI) [23, 30] as two body residual interaction to get self-consistent wave functions. This is a reasonable interaction in this model space and has been used in many studies in this mass region [27, 31–34]. The residual interaction causes mixing of single-particle orbits of nucleons, leading to an average deformed field (Hartree-Fock field). The deformed orbits are generated by the self consistent iterative of the Hartree-Fock equations. Here we use an axially symmetric basis with K quantum number for each intrinsic state. This is actually not a limitation of our model, because we can diagonalise among various K configurations after projection.

Let $|\eta m\rangle$ be a deformed orbit which can be written as:

$$|\eta m\rangle = \sum_j C_{jm}|jm\rangle \tag{1}$$

where j is the angular momentum of the spherical single particle state and m its projection. The amplitude $C_{jm}$ of the spherical state $|jm\rangle$ in the orbit $\eta$ obtained by solving deformed Hartree-Fock equations

$$E^\eta C_{jm}^\eta = \epsilon_j C_{jm}^\eta + \sum (j_3 m_3 j_4 m_4 |V| j m j_2 m_2)\rho_{j_2 m_2 j_4 m_4} C_{j_3 m_3}^\eta \tag{2}$$

where $\epsilon_j$ is the single particle energy of shell model orbit $j(\equiv nlj)$, V denotes the two-body interaction among nucleons and the density matrix

$$\rho_{j_2 m_2 j_4 m_4} = \langle \phi_K|a_{j_4 m_4}^\dagger a_{j_2 m_2}|\phi_K\rangle = \sum_{\eta(occupied)} C_{j_4 m_4}^{*\eta} C_{j_2 m_2}^\eta \tag{3}$$

For axial symmetry one has the conditions $m_3 = m$ and $m_4 = m_2$. Equations (2) and (3) are solved iteratively till the convergence obtained. The state $|\eta m\rangle$ for protons and neutrons form the deformed orbits. The Slater determinants for protons and neutrons constitute the deformed intrinsic state $|\phi_K\rangle$.

### B. Angular Momentum (J) Projection

Because of mixing in single particle orbits, intrinsic states and the multi nucleons Slater determinants are not good angular momentum (J) states (these are states of good K because of assumption of axial symmetries). Thus $|\phi_K\rangle$ does not have a unique J quantum number and is superposition of various J states:

$$|\phi_K\rangle = \sum_J C_K^J|\psi_{JK}\rangle \tag{4}$$

To obtain the band structure and other electromagnetic properties one needs good angular momentum (J) states. The good-J states are obtained by J-projection from deformed intrinsic configurations. This is an important aspect of our microscopic theory. Angular momentum projection from various intrinsic states gives the bands in a nucleus. The angular momentum projection operator is [35]:

$$P_K^{JM} = \frac{2J+1}{8\pi^2} \int d\Omega \; D_{MK}^{J}{}^{*}(\Omega) \; R(\Omega) \tag{5}$$

Here $R(\Omega)$ is the rotation operator $e^{-i\alpha J_z}e^{-i\beta J_y}e^{-i\gamma J_z}$ and $\Omega$ represents the Euler angles $\alpha$, $\beta$ and $\gamma$. For details of HF and angular momentum projection see Refs. [22, 23, 36]. The normalized state of good angular momentum (J) projected from intrinsic state $|\phi_K\rangle$ can be given as

$$|\psi_K^{JM}\rangle = \frac{P_K^{JM}|\phi_K\rangle}{\sqrt{\langle\phi_K|P_K^{JM}|\phi_K\rangle}} \tag{6}$$

The matrix element of the nuclear Hamiltonian, containing single-particle and two-body residual interaction term, between two states of angular momentum J projected from intrinsic states $\phi_{K_1}$ and $\phi_{K_2}$ is:

$$H_{K_1 K_2}^{J} = \frac{2J+1}{2} \frac{1}{(N_{K_1 K_1}^{J} N_{K_2 K_2}^{J})^{1/2}} \int_0^{\pi} d\beta \; sin(\beta) d_{K_1 K_2}^{J}(\beta) \langle\phi_{K_1}|He^{-i\beta J_y}|\phi_{K_2}\rangle \tag{7}$$

where

$$N_{K_1 K_2}^{J} = \frac{2J+1}{2} \int_0^{\pi} d\beta \; sin(\beta) d_{K_1 K_2}^{J}(\beta) \langle\phi_{K_1}|e^{-i\beta J_y}|\phi_{K_2}\rangle \tag{8}$$

Euler angles $\alpha$ and $\gamma$ are integrated out because of axial symmetry and the remaining Euler angle $\beta$ has to be integrated numerically.

Reduced matrix elements of tensor operator $T^L$ between projected states $\psi_{K1}^{J_1}$ and $\psi_{K2}^{J_2}$ are given by

$$\langle\psi_{K_1}^{J_1}||T^L||\psi_{K_2}^{J_2}\rangle = \frac{1}{2} \frac{(2J_2+1)(2J_1+1)^{1/2}}{(N_{K_1 K_1}^{J_1} N_{K_2 K_2}^{J_2})^{1/2}} \sum_{\mu\nu} C_{\mu\nu K_1}^{J_2 L J_1} \int_0^{\pi} d\beta \; sin(\beta) d_{\mu K_2}^{J_2}(\beta) \langle\phi_{K_1}|T_\nu^L e^{-i\beta J_y}|\phi_{K_2}\rangle \tag{9}$$

where the tensor operator $T^L$ denotes electromagnetic operators of multipolarity L.

### C. Bandmixing

In general, two states $|\psi_{K_1}^{JM}\rangle$ and $|\psi_{K_2}^{JM}\rangle$ projected from two intrinsic configurations $|\phi_{K_1}\rangle$ and $|\phi_{K_2}\rangle$ are not orthogonal to each other even if the intrinsic states $|\phi_{K_1}\rangle$ and $|\phi_{K_2}\rangle$ are orthogonal. We orthonormalise them using following equation

$$\sum_{K'} (H_{KK'}^{J} - E_J N_{KK'}^{J}) b_{K'}^{J} = 0 \tag{10}$$

Here $N_{KK'}^{J}$ are amplitude overlap and $b_{K'}^{J}$ are the orthonormalised amplitudes, which can be identified as bandmixing amplitudes. The orthonormalised states are given by

$$\Psi^{JM} = \sum_K b_K^J \psi_K^{JM} \tag{11}$$

With these orthonormalised states we can calculate matrix elements of various tensor operators.

## III. RESULTS AND DISCUSSION

### A. The deformed single particle configurations

The deformed HF orbits are calculated with a spherical core of $^{132}$Sn, the model space spans the $2s_{1/2}$, $1d_{3/2}$, $1d_{5/2}$, $0g_{7/2}$, $0h_{9/2}$, $0h_{11/2}$ orbits for protons and the $2p_{1/2}$, $2p_{3/2}$, $1f_{5/2}$, $1f_{7/2}$, $0h_{9/2}$, $0i_{13/2}$ orbits for neutrons. Single particle energies of these orbits used for both HF and J projection calculations are given in the Table I. We use surface delta interaction as the residual interaction among the active nucleons in these orbits. The interaction has the following form (see Ref. [30])

$$
\begin{aligned}
V(r_{12}) &= -2F(R_0 u_0)^{-4}\delta(r_1 - R_0)\delta(r_2 - R_0)\delta(\cos\omega_{12} - 1) \\
&= -V_0 \sum_{lm} Y_{lm}{}^*(\omega_1)Y_{lm}(\omega_2)
\end{aligned} \tag{12}
$$

where $R_0$ is the nuclear radius, $u_0$ is the radial wave function. Here the radial wave functions are approximated to be same at the nuclear surface. $V_0$ is the strength of the interaction and is taken to be 0.3 MeV for proton-proton (pp), neutron-neutron (nn) and proton-neutron (pn) interactions in the present calculation. This interaction gives a good description of the systematics of deformations in this mass region [31, 32].

TABLE I: Single Particle Energies of the model space

| Proton | $s_{1/2}$ | $d_{3/2}$ | $d_{5/2}$ | $g_{7/2}$ | $h_{9/2}$ | $h_{11/2}$ |
|--------|-----------|-----------|-----------|-----------|-----------|------------|
| [MeV]  | 3.654     | 3.288     | 0.731     | 0         | 7.1       | 2.305      |

| Neutron | $p_{1/2}$ | $p_{3/2}$ | $f_{5/2}$ | $f_{7/2}$ | $h_{9/2}$ | $i_{13/2}$ |
|---------|-----------|-----------|-----------|-----------|-----------|------------|
| [MeV]   | 4.462     | 2.974     | 3.432     | 0         | 0.300     | 1.487      |

The Hartree-Fock solutions for even-even $^{148-160}$Nd nuclei are given in Table II. Since the prolate shape is lower in energy than the oblate for these nuclei, we have used the prolate deformed Hartree-Fock solutions of $^{148-160}$Nd for studying the band structures. As an example, the prolate HF orbits for $^{152}$Nd are shown in Fig. 1.

TABLE II: Hartree-Fock solutions for $_{60}$Nd nuclei.

| A | Shape | $E_{HF}$ (MeV) | $\langle Q_{20} \rangle$ Proton | Neutron | $\beta$ Theory | Experiment [11] |
|---|-------|------|---------|---------|--------|-----------------|
| 148 | Prolate | -45.195 | 12.275 | 12.676 | 0.273 | 0.201±0.004 |
|     | Oblate | -42.875 | -10.745 | -10.885 | -0.237 | |
| 150 | Prolate | -56.163 | 12.332 | 15.030 | 0.287 | 0.285±0.002 |
|     | Oblate | -52.250 | -10.962 | -12.080 | -0.247 | |
| 152 | Prolate | -67.303 | 12.387 | 17.113 | 0.300 | 0.349±0.012 |
|     | Oblate | -66.679 | -11.139 | -14.614 | -0.265 | |
| 154 | Prolate | -78.483 | 12.600 | 18.006 | 0.308 | |
|     | Oblate | -61.842 | -4.178 | -8.337 | -0.117 | |
| 156 | Prolate | -91.123 | 12.631 | 20.089 | 0.320 | |
|     | Oblate | -82.592 | -9.727 | -16.885 | -0.256 | |
| 158 | Prolate | -102.957 | 12.660 | 21.614 | 0.329 | |
|     | Oblate | -101.640 | -11.503 | -19.432 | -0.297 | |
| 160 | Prolate | -113.350 | 12.739 | 23.196 | 0.338 | |
|     | Oblate | -112.814 | -11.574 | -19.747 | -0.299 | |

The quadrupole deformation parameter ($\beta$) values obtained in the HF calculations are also listed in Table II. We have used the effective charges of 1.7e and 0.7e for protons and neutrons respectively. The experimental trend of $\beta$ values (known for A=148-152) are reproduced quite well. The $\beta$ values show an increasing trend till $A = 160$.

Because of time-reversal symmetry the nucleon orbits $|\pm\Omega\rangle$ are doubly degenerate for these even-even nuclei. We make suitable neutrons and protons excitation across the Fermi surfaces and considered suitable intrinsic configurations for angular momentum projection. These intrinsic states are listed in Table III.

As one of our main interests is to study low-lying K-isomers, so it is important to give a closer look to the levels near the proton and neutron Fermi surfaces. Protons orbits based on $h_{11/2}(1/2)$ and $h_{11/2}(3/2)$ are just below the proton Fermi surface and orbits based on $g_{7/2}(5/2)$ and $h_{11/2}(5/2)$ are just above it. Hence, it is expected that these orbits can be coupled to give $K = 4$ and $K = 5$ isomeric states. In neutron side, the orbits near neutron Fermi surface are based on $f_{7/2}$, $h_{9/2}$ and $i_{13/2}$. The K configurations from proton side is uniform but as the neutron number changes through out the isotopic chain there are considerable variations around the neutron Fermi surfaces leading to the change in the configurations of K-isomers built on quasi-neutrons.

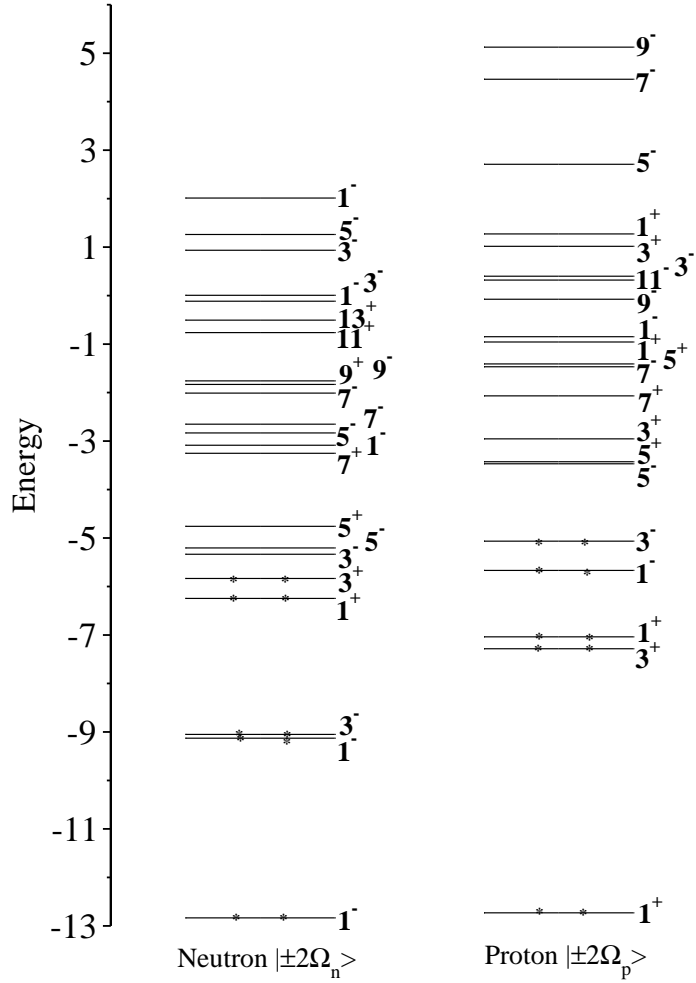

FIG. 1: Prolate deformed HF orbits of $^{152}$Nd. The orbits are denoted by $|2m|^\pi$. Occupied orbits are marked by (*).

TABLE III: Intrinsic configurations for J-projection calculation for $_{60}$Nd nuclei.

| Isotope | Index | $K^\pi$ | Configuration |
|---------|-------|---------|---------------|
| 148 | B1 | $0^+$ | HF |
| | B2 | $1^-$ | $\pi(3/2^- \otimes 5/2^+)$ |
| | B3 | $0^-$ | $\pi(3/2^- \otimes 3/2^+)$ |
| | B4 | $2^-$ | $\pi(1/2^- \otimes 5/2^+)$ |
| | B5 | $4^-$ | $\pi(3/2^- \otimes 5/2^+)$ |
| | B6 | $5^-$ | $\pi(5/2^- \otimes 5/2^+)$ |
| | B7 | $5^-$ | $\nu(5/2^- \otimes 5/2^+)$ |
| | B8 | $9^+$ | $\pi(3/2^- \otimes 5/2^+) \otimes \nu(5/2^- \otimes 5/2^+)$ |
| | B9 | $10^+$ | $\pi(5/2^- \otimes 5/2^+) \otimes \nu(5/2^- \otimes 5/2^+)$ |
| 150 | B1 | $0^+$ | HF |
| | B2 | $1^-$ | $\pi(3/2^- \otimes 5/2^+)$ |
| | B3 | $4^-$ | $\pi(3/2^- \otimes 5/2^+)$ |
| | B4 | $5^-$ | $\pi(5/2^- \otimes 5/2^+)$ |
| | B5 | $5^-$ | $\nu(5/2^- \otimes 5/2^+)$ |

TABLE III – Continued

| Isotope | Index | $K^\pi$ | Configuration |
|---------|-------|---------|---------------|
| | B6 | $9^+$ | $\pi(3/2^- \otimes 5/2^+) \otimes \nu(5/2^- \otimes 5/2^+)$ |
| | B7 | $10^+$ | $\pi(5/2^- \otimes 5/2^+) \otimes \nu(5/2^- \otimes 5/2^+)$ |
| 152 | B1 | $0^+$ | HF |
| | B2 | $0^-$ | $\nu(3/2^+ \otimes 3/2^-)$ |
| | B3 | $1^-$ | $\nu(3/2^+ \otimes 5/2^-)$ |
| | B4 | $1^-$ | $\nu(3/2^- \otimes 5/2^+)$ |
| | B5 | $2^-$ | $\nu(1/2^+ \otimes 5/2^-)$ |
| | B6 | $4^-$ | $\nu(3/2^- \otimes 5/2^+)$ |
| | B7 | $4^-$ | $\pi(3/2^- \otimes 5/2^+)$ |
| | B8 | $5^-$ | $\pi(5/2^- \otimes 5/2^+)$ |
| | B9 | $5^-$ | $\nu(5/2^- \otimes 5/2^+)$ |
| | B10 | $9^+$ | $\pi(3/2^- \otimes 5/2^+) \otimes \nu(5/2^- \otimes 5/2^+)$ |
| | B11 | $10^+$ | $\pi(5/2^- \otimes 5/2^+) \otimes \nu(5/2^- \otimes 5/2^+)$ |
| 154 | B1 | $0^+$ | HF |
| | B2 | $1^-$ | $\pi(3/2^- \otimes 5/2^+)$ |
| | B3 | $1^-$ | $\nu(3/2^- \otimes 5/2^+)$ |
| | B4 | $4^-$ | $\pi(3/2^- \otimes 5/2^+)$ |
| | B5 | $4^-$ | $\nu(3/2^- \otimes 5/2^+)$ |
| | B6 | $5^-$ | $\pi(5/2^- \otimes 5/2^+)$ |
| | B7 | $5^-$ | $\nu(5/2^- \otimes 5/2^+)$ |
| | B8 | $9^+$ | $\pi(3/2^- \otimes 5/2^+) \otimes \nu(5/2^- \otimes 5/2^+)$ |
| | B9 | $10^+$ | $\pi(5/2^- \otimes 5/2^+) \otimes \nu(5/2^- \otimes 5/2^+)$ |
| 156 | B1 | $0^+$ | HF |
| | B2 | $1^-$ | $\pi(3/2^- \otimes 5/2^+)$ |
| | B3 | $4^-$ | $\pi(3/2^- \otimes 5/2^+)$ |
| | B4 | $5^-$ | $\pi(5/2^- \otimes 5/2^+)$ |
| | B5 | $5^-$ | $\nu(5/2^- \otimes 5/2^+)$ |
| | B6 | $9^+$ | $\pi(3/2^- \otimes 5/2^+) \otimes \nu(5/2^- \otimes 5/2^+)$ |
| | B7 | $10^+$ | $\pi(5/2^- \otimes 5/2^+) \otimes \nu(5/2^- \otimes 5/2^+)$ |
| 158 | B1 | $0^+$ | HF |
| | B2 | $1^-$ | $\pi(3/2^- \otimes 5/2^+)$ |
| | B3 | $4^-$ | $\pi(3/2^- \otimes 5/2^+)$ |
| | B4 | $5^-$ | $\pi(5/2^- \otimes 5/2^+)$ |
| | B5 | $5^-$ | $\nu(5/2^- \otimes 5/2^+)$ |
| | B6 | $9^+$ | $\pi(3/2^- \otimes 5/2^+) \otimes \nu(5/2^- \otimes 5/2^+)$ |
| | B7 | $10^+$ | $\pi(5/2^- \otimes 5/2^+) \otimes \nu(5/2^- \otimes 5/2^+)$ |
| 160 | B1 | $0^+$ | HF |
| | B2 | $1^-$ | $\pi(3/2^- \otimes 5/2^+)$ |
| | B3 | $4^-$ | $\pi(3/2^- \otimes 5/2^+)$ |
| | B4 | $5^-$ | $\pi(5/2^- \otimes 5/2^+)$ |
| | B5 | $4^-$ | $\nu(1/2^- \otimes 7/2^+)$ |
| | B6 | $8^+$ | $\pi(3/2^- \otimes 5/2^+) \otimes \nu(1/2^- \otimes 7/2^+)$ |
| | B7 | $9^+$ | $\pi(5/2^- \otimes 5/2^+) \otimes \nu(1/2^- \otimes 7/2^+)$ |

## B.  Energy Spectra

Deformed Hartree-Fock and Angular Momentum Projection calculations are performed for even-even Nd isotopic chain with mass number $A = 148$ to $A = 160$. The energy spectra associated with each intrinsic state is obtained by angular momentum projection in the given model space and with surface delta residual interaction, using the formalism discussed in Section II. The energy spectra, band-head energies and other spectroscopic properties are shown in Tables IV-V and in Figures 2-14 for even-even $^{148-160}$Nd isotopes.

TABLE IV: Comparison of calculated and experimentally observed transition energies , reduced transition probabilities $B(E2:J \to J-2)$, spectroscopic quadrupole moments $Q_S$ and $g$-factors $g(J)$ of $_{60}$Nd nuclei. Here B(E2) and Q(J) are calculated for effective charges $e_p = 1 + e_{eff}$ and $e_n = e_{eff}$ where $e_{eff} = 0.7$. The g-factors of $g_l^{\pi} = 1.0$, $g_l^{\nu} = 0.0$, $g_s^{\pi} = 5.586 \times 0.5$ and $g_s^{\nu} = -3.826 \times 0.5$ are used to calculate $g(J)$. The experimental transition energies are taken from Ref. [7] for $^{148,150}$Nd, Ref. [8] for $^{152}$Nd and Ref. [9] for $^{154,156}$Nd. The experimental B(E2) and $Q_S$ values are from Ref. [11] and that of $g(J)$ are from Ref. [12]

| Transition | Transition Energy (MeV) | | $B(E2:J \to J-2)$ ($e^2$b$^2$) | | $Q_S$ (eb) | | $g(J)$ (nm) | |
|---|---|---|---|---|---|---|---|---|
| $J \to J-2$ | Theory | Experiment [Ref.] | Theory | Experiment | Theory | Experiment | Theory | Experiment |
| $^{148}$Nd | | | | | | | | |
| $2^+ \to 0^+$ | 0.1122 | 0.302 | 0.5228 | 0.276±0.004 | -1.4643 | -1.46± 0.13 | 0.6153 | 0.365±0.015 |
| $4^+ \to 2^+$ | 0.2603 | 0.451 | 0.7440 | 0.428±0.004 | -1.8624 | – | 0.6157 | 0.35±0.05 |
| $6^+ \to 4^+$ | 0.4047 | 0.527 | 0.8136 | – | -2.0465 | – | 0.6162 | 0.267±0.050 |
| $8^+ \to 6^+$ | 0.5438 | 0.576 | 0.8430 | – | -2.1502 | – | 0.6169 | – |
| $10^+ \to 8^+$ | 0.6765 | 0.615 | 0.8539 | – | -2.2163 | – | 0.6181 | -0.175±0.009 |
| $12^+ \to 10^+$ | 0.8018 | 0.635 | 0.8545 | – | -2.2599 | – | 0.6192 | – |
| $^{150}$Nd | | | | | | | | |
| $2^+ \to 0^+$ | 0.0719 | 0.130 | 0.5794 | 0.542±0.006 | -1.5413 | -2.0± 0.5 | 0.3828 | 0.42±0.02 |
| $4^+ \to 2^+$ | 0.1646 | 0.251 | 0.8265 | 0.820±0.036 | -1.9587 | – | 0.3808 | 0.44±0.03 |
| $6^+ \to 4^+$ | 0.2508 | 0.339 | 0.9074 | 0.991±0.055 | -2.1506 | – | 0.3777 | 0.35±0.07 |
| $8^+ \to 6^+$ | 0.3278 | 0.409 | 0.9453 | 1.136±0.02 | -2.2592 | – | 0.3740 | 0.56±0.13 |
| $10^+ \to 8^+$ | 0.3951 | 0.470 | 0.9656 | 0.996±0.02 | -2.3275 | – | 0.3701 | 0.14±0.19 |
| $12^+ \to 10^+$ | 0.4541 | 0.520 | 0.9754 | – | -2.3742 | – | 0.3662 | – |
| $^{152}$Nd | | | | | | | | |
| $2^+ \to 0^+$ | 0.0725 | 0.072 | 0.6432 | 0.84±0.06 | -1.6241 | – | 0.3755 | – |
| $4^+ \to 2^+$ | 0.1683 | 0.164 | 0.9169 | 1.04±0.05 | -2.0657 | – | 0.3754 | – |
| $6^+ \to 4^+$ | 0.2619 | 0.247 | 1.0069 | 1.06±0.02 | -2.2705 | – | 0.3755 | – |
| $8^+ \to 6^+$ | 0.3526 | 0.322 | 1.0490 | – | -2.3876 | – | 0.3755 | – |
| $10^+ \to 8^+$ | 0.4395 | 0.389 | 1.0709 | – | -2.4619 | – | 0.3755 | – |
| $12^+ \to 10^+$ | 0.5220 | 0.452 | 1.0817 | – | -2.5124 | – | 0.3756 | – |
| $^{154}$Nd | | | | | | | | |
| $2^+ \to 0^+$ | 0.0713 | 0.071 | 0.6888 | 0.45±0.12 | -1.6797 | – | 0.3822 | – |
| $4^+ \to 2^+$ | 0.1633 | 0.162 | 0.9825 | – | -2.1355 | – | 0.3801 | – |
| $6^+ \to 4^+$ | 0.2485 | 0.248 | 1.0792 | – | -2.3446 | – | 0.3769 | – |
| $8^+ \to 6^+$ | 0.3243 | 0.328 | 1.1253 | – | -2.4627 | – | 0.3729 | – |
| $10^+ \to 8^+$ | 0.3904 | 0.400 | 1.1499 | – | -2.5367 | – | 0.3688 | – |
| $12^+ \to 10^+$ | 0.4482 | 0.466 | 1.1631 | – | -2.5871 | – | 0.3649 | – |
| $^{156}$Nd | | | | | | | | |
| $2^+ \to 0^+$ | 0.0701 | 0.067 | 0.7571 | – | -1.7615 | – | 0.3740 | – |
| $4^+ \to 2^+$ | 0.1626 | 0.155 | 1.0800 | – | -2.2409 | – | 0.3739 | – |
| $6^+ \to 4^+$ | 0.2531 | 0.239 | 1.1863 | – | -2.4628 | – | 0.3738 | – |
| $8^+ \to 6^+$ | 0.3407 | 0.318 | 1.2366 | – | -2.5893 | – | 0.3737 | – |
| $10^+ \to 8^+$ | 0.4245 | 0.391 | 1.2636 | – | -2.6691 | – | 0.3736 | – |
| $12^+ \to 10^+$ | 0.5043 | 0.459 | 1.2772 | – | -2.7231 | – | 0.3734 | – |
| $^{158}$Nd | | | | | | | | |
| $2^+ \to 0^+$ | 0.0779 | – | 0.8146 | – | -1.8277 | – | 0.3936 | – |
| $4^+ \to 2^+$ | 0.1810 | – | 1.1618 | – | -2.3245 | – | 0.3936 | – |
| $6^+ \to 4^+$ | 0.2823 | – | 1.2758 | – | -2.5538 | – | 0.3935 | – |
| $8^+ \to 6^+$ | 0.3810 | – | 1.3295 | – | -2.6845 | – | 0.3934 | – |
| $10^+ \to 8^+$ | 0.4763 | – | 1.3578 | – | -2.7670 | – | 0.3935 | – |
| $12^+ \to 10^+$ | 0.5676 | – | 1.3722 | – | -2.8223 | – | 0.3934 | – |
| $^{160}$Nd | | | | | | | | |
| $2^+ \to 0^+$ | 0.0734 | – | 0.8776 | – | -1.8974 | – | 0.3774 | – |
| $4^+ \to 2^+$ | 0.1706 | – | 1.2516 | – | -2.4130 | – | 0.3775 | – |
| $6^+ \to 4^+$ | 0.2664 | – | 1.3749 | – | -2.6525 | – | 0.3774 | – |

TABLE IV – Continued

| Transition | Transition Energy (MeV) | | | $B(E2:J \rightarrow J-2)$ $(e^2b^2)$ | | $Q_S$ (eb) | | $g(J)$ (nm) | |
|---|---|---|---|---|---|---|---|---|---|
| $J \rightarrow J-2$ | Theory | Experiment | [Ref.] | Theory | Experiment | Theory | Experiment | Theory | Experiment |
| $8^+ \rightarrow 6^+$ | 0.3601 | – | | 1.4327 | – | -2.7893 | – | 0.3775 | – |
| $10^+ \rightarrow 8^+$ | 0.4510 | – | | 1.4634 | – | -2.8761 | – | 0.3775 | – |
| $12^+ \rightarrow 10^+$ | 0.5387 | – | | 1.4791 | – | -2.9349 | – | 0.3775 | – |

### 1. *The Ground Bands*

The $K = 0^+$ ground bands of even-even $^{148-160}$Nd have rotational behaviour with the J values extending upto $32\hbar$ in our calculations (we have shown the ground band upto about $26\hbar$ because of limitations of energy scale) with the excitation energy about 10 MeV. Experimentally the $K = 0^+$ ground bands of even-even $^{148-156}$Nd nuclei are known upto about $16\hbar$ and they are quite well described by our theoretical work (see Figures 2-6). The transition energies of few lowest states of these nuclei in comparison with available experimental data are compared in Table IV. For $^{158}$Nd and $^{160}$Nd energy spectra are not known experimentally and we predict a good rotational band structures for these nuclei as shown in Figures 7 and 8 (the bands in these figures are all predictions for these two neutron-rich nuclei).

### 2. *The Negative Parity Bands*

The nucleus $^{148}$Nd exhibits a low-lying negative parity band with $3^-$ state lying below $1^-$ in an early beta decay experiment by Walters et al. [37] and later it was extended by Ibbotson et al. [38]. Our band mixing (by mixing B2, B3 and B4 as given in Table III for $^{148}$Nd) calculation reasonably reproduces the spectrum of this band with $3^-$ state lower in energy than $1^-$ state as in experiment. The lowest band after band mixing calculation is compared with the experimentally observed band and shown in Fig. 2. The bandhead energies of the calculated $K = 1^-$ and $K = 2^-$ negative parity bands are listed in Table V.

In Fig. 3, the calculated $K = 1^-$ band for $^{150}$Nd is compared with the experimentally known negative parity band [39]. Though the calculated bandhead energy is about 400 keV higher than that of the experimental value, good agreement of moment of inertia between the calculated and experimental bands help us to conclude that the negative parity band could be of $K = 1^-$ ($\pi 3/2^-[541] \otimes \pi 5/2^+[413]$) origin as predicted in Ref. [32]. We have also predicted the unfavoured even-J branch of this band. The intensity to populate these unfavoured states may be small, but still can be looked for with multi-detector arrays. The signature splitting in this band is evident from the doublet structure because of the rotational alignment of $\pi h_{11/2}$.

After mixing a few low energy intrinsic configurations (B2, B3 and B4) for $^{152}$Nd we obtained negative parity bands. The calculated negative parity bands are compared with experimental $K = 0^-$ and $K = 1^-$ bands (see Fig. 4). The signature splitting is observed in our calculation for both the bands. For $K = 0^-$ band only lowest 3 levels of odd-J branch are experimentally known [40] and these are compared with our calculated favoured branch. The other branch is not observed in experiment. From an early experiment Hellström et al. [40] find a $K = 2^-$ band but recently Yeoh et al. [29] assigned this as a $K = 1^-$ band with configuration $\nu 5/2^+[642] \otimes \nu 3/2^-[521]$. Experimentally this band is known upto $12\hbar$ (only even J states). We have compared our calculated $K = 1^-$ band with this in Figure 4.

A similar $K = 1^-$ band is also known experimentally for $^{154}$Nd [9]. The calculated band in comparison with available experimental data is shown in Fig. 5. An excellent agreement is obtained between calculated and experiment spectra for this isotope. No $K = 1^-$ band has been observed experimentally for $^{156,158,160}$Nd. Our predictions for these bands are shown in Figures 6-8.

### 3. *The K-Isomer Bands*

Experimentally a low lying $K = 5^-$ band (with a few states built on the bandhead) is known in $^{156}$Nd [9], while for $^{154}$Nd only a single $K = 5^-$ isomeric state in known [10]. We find low lying $K = 5^-$ bands for all Nd nuclei studied in this article (see configurations in Table III). In our PHF calculation the $K = 5^-$ proton 2-qp band is lower in energy by about 5 MeV than the $K = 5^-$ neutron 2-qp band for $^{148}$Nd (see Fig. 2). In all other cases (for $A = 150 - 160$) the neutron 2-qp bands are lower in energy. There is a large energy gap between the filled and next unoccupied orbit on the neutron side in ground state configuration of $^{148}$Nd. Therefore it requires more energy to excite particles across the neutron Fermi surface. While for the other isotopes the orbits are closed to the Fermi level so low energy neutron excitations are possible. There is no experimental evidence for the K-isomer band in $^{150}$Nd. Our predictions for 2qp K-isomer band for $^{150}$Nd and $^{152}$Nd are shown in Fig. 3 and Fig. 4 respectively. The calculated $K = 5^-$ isomeric band for $^{154}$Nd is given in Fig. 5 and compared with experimental findings [10]. Our predicted $K = 5^-$ band-head is about 300 keV higher than experimental value. Similar $K = 5^-$ band is recently observed for $^{156}$Nd [9] and this band is experimentally known upto $13\hbar$. We predict reasonably this band and our calculated band is extended upto $32\hbar$. Apart from these we have also studied $K = 4^-$ 2-qp bands from protons/neutrons excitations and these are listed in Table V. Recently, $K = 4^-$ isomer band is observed experimentally for $^{154}$Nd [9] with dominant configuration as $\nu 5/2^+[642] \otimes \nu 3/2^-[521]$. Our PHF calculations also give a neutron 2-qp band with same configuration. Similar band is also obtained in recent Projected Shell Model (PSM) calculations [41]. As it can be seen from the Table V that proton 2-qp $K = 4^-$ bands are low lying with bandhead energies less than 2 MeV (except for $^{152}$Nd it is about 2 MeV). For Nd isotopes with $N > 98$ the $i_{\frac{13}{2}}(7/2)$ state is very close to the neutron Fermi surface so we obtained a low-lying neutron 2-qp $K = 4^-$ band with bandhead energy about 0.5 MeV for $^{160}$Nd.

Apart form the 2-qp bands we have also studied the 4-qp $K = 8^+$, $K = 9^+$ and $K = 10^+$ bands obtained from the combination of proton and neutron 2-qp excitations. Our predictions for these bands are shown in Figures 2-8. All these bands are identical (bandhead energies and electromagnetic properties are shown in Tab. V) for the nuclei studied here and for neutron rich nuclei these high-K isomer bands are low-lying. The configurations for these bands are also listed in Table III.

TABLE V: Comparison of calculated and experimentally observed band head energy (BHE), spectroscopic quadrupole moments ($Q_S$) and magnetic moment ($\mu$) of Nd nuclei. Here $Q_S$ are calculated with effective charges $e_p = 1 + e_{eff}$ and $e_n = e_{eff}$ where $e_{eff} = 0.7$. The g-factors of $g_l^\pi = 1.0$, $g_l^\nu = 0.0$, $g_s^\pi = 5.586 \times 0.5$ and $g_s^\nu = -3.826 \times 0.5$ are used to calculate $\mu$.

| Nuclei | $K^\pi$ | BHE (MeV) | | $Q_S$ (eb) | $\mu$ (nm) |
|---|---|---|---|---|---|
| | | Theory | Experiment [Ref.] | | |
| $^{148}$Nd | $1^-$(a)§ | 1.1584 | 0.9992 [38] | -1.2078 | 1.7655 |
| | $2^-$(a) | 1.8100 | – | 1.2936 | 2.0305 |
| | $4^-$(a) | 1.6189 | – | 2.4688 | 3.6675 |
| | $5^-$(a) | 3.2242 | – | 2.6836 | 4.7003 |
| | $5^-$(b) | 8.3077 | – | 2.7182 | 0.2112 |
| | $9^+$ | 9.1619 | – | 3.2392 | 3.8401 |
| | $10^+$ | 10.2181 | – | 3.2406 | 4.8406 |
| | | | | | |
| $^{150}$Nd | $1^-$(a) | 1.2541 | 0.8529 [39] | 0.5138 | 0.3045 |
| | $4^-$(a) | 1.6754 | – | 2.6097 | 3.5146 |
| | $5^-$(b) | 2.6951 | – | 3.0178 | 0.2163 |
| | $5^-$(a) | 3.1465 | – | 2.8669 | 4.5417 |
| | $9^+$ | 4.1121 | – | 3.6184 | 3.8383 |
| | $10^+$ | 5.1456 | – | 3.6343 | 4.8607 |
| | | | | | |
| $^{152}$Nd | $0^-$(b)§ | 1.1947 | 1.1486 [40] | -0.9279 | 0.3723 |
| | $1^-$(b) | 1.4778 | – | 0.5629 | 0.2091 |
| | $4^-$(a) | 2.0516 | – | 2.7567 | 3.4989 |
| | $4^-$(b) | 1.1094 | – | 2.6080 | 0.1386 |
| | $5^-$(b) | 2.3347 | – | 2.2677 | 0.1347 |
| | $5^-$(a) | 3.7741 | – | 3.0344 | 4.5384 |
| | $9^+$ | 3.9071 | – | 3.8367 | 3.7329 |
| | $10^+$ | 5.0516 | – | 3.8698 | 4.7339 |
| | | | | | |
| $^{154}$Nd | $1^-$(b) | 0.9412 | 0.9619 [9] | 0.5625 | 0.2993 |
| | $4^-$(a) | 2.8622 | – | 2.8575 | 3.5161 |
| | $4^-$(b) | 1.0357 | 1.2980 [9] | 3.0174 | 0.2687 |
| | $5^-$(b) | 1.6644 | 1.348 [10] | 3.3648 | 0.1306 |
| | $5^-$(a) | 3.1832 | – | 3.1466 | 4.5386 |
| | $9^+$ | 3.0651 | – | 4.0577 | 3.7488 |
| | $10^+$ | 4.0333 | – | 4.0969 | 4.7390 |
| | | | | | |
| $^{156}$Nd | $1^-$(a) | 1.8127 | – | 3.0021 | 3.5005 |
| | $4^-$(a) | 1.7308 | – | 3.0021 | 3.5005 |
| | $5^-$(b) | 1.2186 | 1.4312 [9] | 3.5664 | 0.2694 |

Continued on next page...

TABLE V – Continued

| Nuclei | K$^\pi$ | BHE (MeV) | | Q$_S$ (eb) | $\mu$ (nm) |
|--------|---------|-----------|------------|------------|------------|
| | | Theory | Experiment [Ref.] | | |
| | 5$^-$(a) | 3.1181 | – | 3.3117 | 4.5353 |
| | 9$^+$ | 3.1211 | – | 4.3061 | 3.8841 |
| | 10$^+$ | 4.4025 | – | 4.3220 | 4.8955 |
| $^{158}$Nd | 1$^-$(a) | 2.0440 | – | 0.6138 | 0.2937 |
| | 4$^-$(a) | 1.9652 | – | 3.1179 | 3.5127 |
| | 5$^-$(b) | 2.8203 | – | 3.6161 | 0.2186 |
| | 5$^-$(a) | 3.3823 | – | 3.4399 | 4.5418 |
| | 9$^+$ | 5.5992 | – | 4.4629 | 3.8729 |
| | 10$^+$ | 6.0933 | – | 4.4437 | 4.8306 |
| $^{160}$Nd | 1$^-$(a) | 1.7625 | – | 0.6379 | 0.3009 |
| | 4$^-$(b) | 0.5272 | – | 3.3536 | 0.2544 |
| | 4$^-$(a) | 1.6855 | – | 3.2417 | 3.5023 |
| | 5$^-$(a) | 2.9253 | – | 3.5807 | 4.5335 |
| | 8$^+$ | 2.2104 | – | 4.4304 | 3.8387 |
| | 9$^+$ | 3.2361 | – | 4.4830 | 4.8425 |

(a) proton excitation (b) neutron excitation
$\S$ lowest state is 3$^-$
$\S$ $K = 0^-$ band, Q$_S$ and $\mu$ are given for 1$^-$ state

## C. Electromagnetic Properties of Bands

### 1. B(E2)

Apart from the energy spectra we have calculated the reduced transition matrix elements e.g. B(E2), B(M1). The B(E2) values for a transition from an initial state $\alpha J_1$ to final state $\beta J_2$ is given by

$$B(E2; \alpha J_1 \rightarrow \beta J_2) = \frac{1}{2J_1 + 1} \left| \sum_{i=p,n} \langle \Psi^{\beta J_2}_{K_2} || Q_2^i || \Psi^{\alpha J_1}_{K_1} \rangle \right|^2 \qquad (13)$$

where the summation is for quadrupole moment operators of active protons and neutrons. The effective charges of 1.7e and 0.7e are used for protons and neutrons respectively in our calculations. The calculated B(E2) values for ground state bands are given in Table IV. Also, in Figures 9 and 10 we have shown the behavior of B(E2) values for ground bands, negative parity bands as well as K-isomer bands for the nuclei studied in this paper. From these figures we find that for the ground band of all the Nd isotopes considered, B(E2) values are smoothly varying with increasing J. The $B(E2; J \rightarrow J-2)$ values for the ground band show similar trend through out the isotopic chain studied here. It is maximum in the range J=12 to J=16 and falling down beyond this range. For other bands in case of $J \rightarrow J-2$ transitions the values are increasing with J and after a certain spin its nearly constant. For the transitions $J \rightarrow J-1$, the maximum B(E2) values are around 1.0 (eb)$^2$ for 2-qp bands and a little bit lower for 4-qp bands. The B(E2) values for $K = 1^-$ bands(proton excitation) are calculated and displayed in Fig. 11. Similar variations in B(E2) values are observed for all the nuclei considered in our calculations. The $B(E2; 2^+ \rightarrow 0^+)$ systematic is compared with experimentally available values. In Figure 12, the experimental trend reasonably reproduced with our calculations.

### 2. B(M1)

The reduced magnetic dipole transition moments between initial and final states are given by

$$B(M1; \alpha J_1 \rightarrow \beta J_2) = \frac{1}{2J_1 + 1} \frac{3}{4\pi} \left| \sum_{i=p,n} \langle \Psi^{\beta J_2}_{K_2} || g_l^i l_i + g_s^i s_i || \Psi^{\alpha J_1}_{K_1} \rangle \right|^2 \qquad (14)$$

where $g_l$ and $g_s$ are orbital and spin g-factor respectively. The g-factors of $g_l = 1.0\mu_N$ and $g_s = 5.586 \times 0.5\mu_N$ for protons and $g_l = 0\mu_N$ and $g_s = -3.826 \times 0.5\mu_N$ for neutrons are used for B(M1) calculations. The quenching of spin g-factors by 0.5 is taken in account to consider the core polarization effect [42, 43].

B(M1) values for K-isomer bands are shown in Fig 13. For $K = 5^-$ proton 2-qp band B(M1) values are gradually increasing with J for all nuclei. B(M1) values for neutron 2-qp band in $^{148,150,152,154}$Nd also show a same behaviour. But for $^{156,158}$Nd , the B(M1) values are constant at higher J. $K = 4^-$ neutron 2-qp band in $^{160}$Nd also shows a similar trend. From Fig. 13, for $^{148}$Nd we can see that at high spin above J=25, both the 2-qp bands show staggering in B(M1) values and it is more prominent in case of neutron 2-qp band because of the doublet structure of these bands at high spin values. For transitions within doublets, the terms contributing to reduce M1 matrix elements are large resulting this staggering effect. For 4-qp bands we notice that the B(M1) values are very small, where the contributions from neutron spin is comparable with that of proton; but as g-factor of neutrons is opposite sign to that of protons, they cancel each other giving a very small net quantity. Similarly for $K = 1^-$ bands (from proton excitations) the B(M1) values are also very small and shown in Fig. 14.

### 3. *Spectroscopic Quadrupole Moment ($Q_S$)*

The spectroscopic quadrupole moment of a state with angular momentum J is given by

$$Q_S(J) = \frac{1}{(2J+1)^{1/2}} \left(\frac{16\pi}{5}\right)^{1/2} C_{J0J}^{J2J} \langle \Psi_K^J || \sum_{i=p,n} Q_2^i || \Psi_K^J \rangle \tag{15}$$

We have calculated the electric quadrupole moments for band head as well as other spin states for all bands. The calculated values are given in Tables IV-V. The experimental values are only known for ground state band and comparisons are given in Table IV. The known experimental results are nicely reproduced in our calculations.

### 4. *Magnetic Dipole Moment ($\mu$) and g-factor*

The magnetic dipole moment ($\mu$) of a state with angular momentum J can be expressed as

$$\mu(J) = \frac{1}{(2J+1)^{1/2}} C_{J0J}^{J1J} \left( \sum_{i=p,n} \langle \Psi_K^J || g_l^i l_i + g_s^i s_i || \Psi_K^J \rangle \right) \tag{16}$$

where i=p and n for protons and neutrons respectively.

The gyromagnetic factor (g-factor) of state J is defined as

$$g(J) = \frac{\mu(J)}{J} \tag{17}$$

where $\mu(J)$ is the magnetic moment of state J.

The g-factors for ground state band for different J values are calculated and are given in Table IV. Experimental data are only available for $^{148,150}$Nd [11]. In our calculations, for $^{148}$Nd the values are overestimated and for $^{150}$Nd, the calculated values are reasonable in comparison to experimental values except for J=10. The magnetic dipole moments for K-isomer and negative parity bands are calculated and given in Table V. No experimental data is available for these bands. The magnetic moments for the $K = 5^-$ neutron 2-qp bands are lower than the proton 2-qp bands.

## IV. CONCLUSION

The neutron-rich even mass Nd nuclei have interesting quadrupole deformation properties and band structures. We have presented results for such nuclei, using the microscopic model of angular momentum projected deformed Hartree-Fock method. Some of our results correlate well with known experimental findings about the deformation and the spectra of these nuclei. Using our PHF model with band mixing we predict many bands (including K-isomer bands) upto high spin values. The spectra for both positive and negative parity, two-quasiparticle and four-quasiparticle bands as well as electromagnetic properties like B(E2), B(M1) of these bands are obtained. The known systematics of the band structures are given quite well with this model. These results can be useful in future experimental studies of these nuclei. As a whole our microscopic model can be very useful to study the band structure of nuclei and electromagnetic properties in this mass region to high spin values.

**Acknowledgments**

SKG acknowledges the financial support from IOP, Bhubaneswer where the part of the work is done. SKP and CRP acknowledge the support of the Department of Science and Technology, India (DST Project SR/S2/HEP-37/2008) during this work. We thanks Prof. R. K. Bhowmik for fruitful discussions.

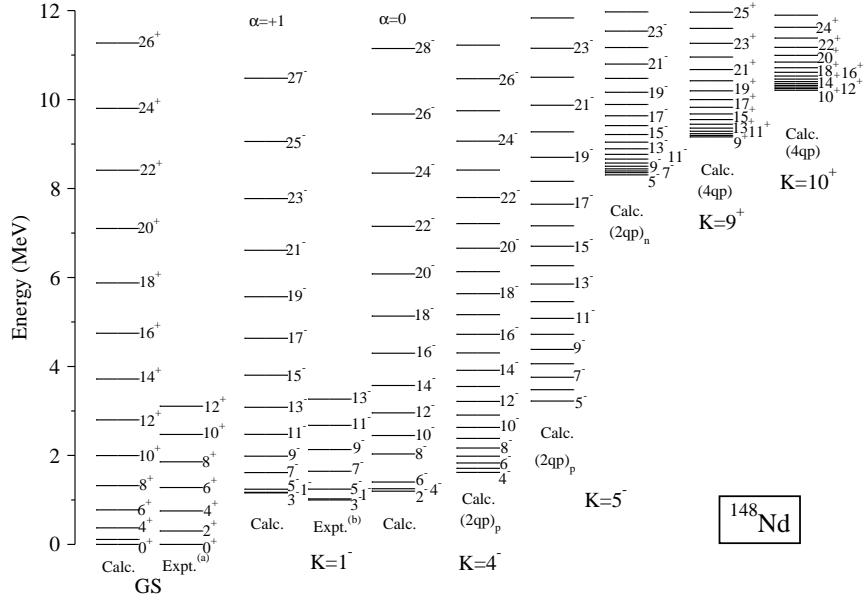

FIG. 2: Comparison of theoretical spectra with experimental spectrum of various band for $^{148}$Nd. (a) Ref. [7], (b) Ref. [38].

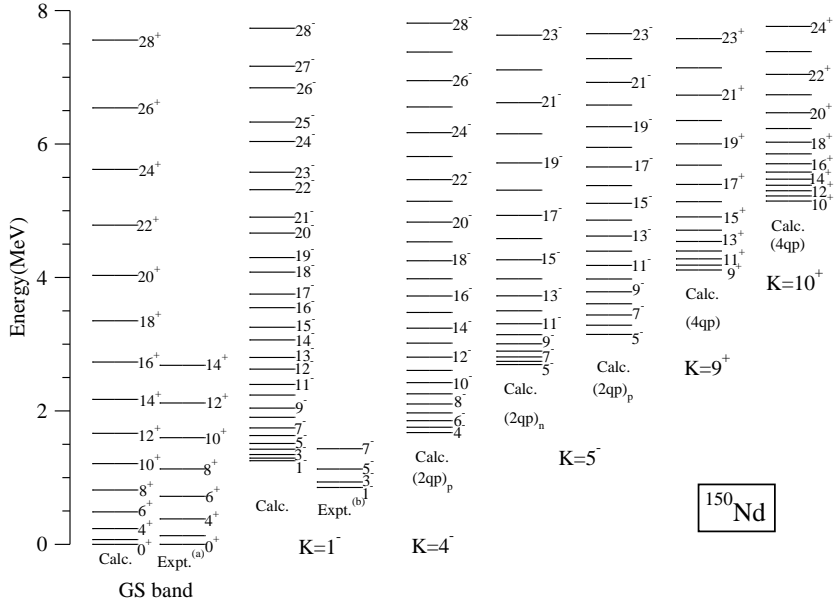

FIG. 3: Same as Fig. 2 but for $^{150}$Nd. (a) Ref. [7], (b) Ref. [39]. The $K = 1^-$ band of our calculation has both odd-J and even-J (unfavoured) states (see discussion in Section III B).

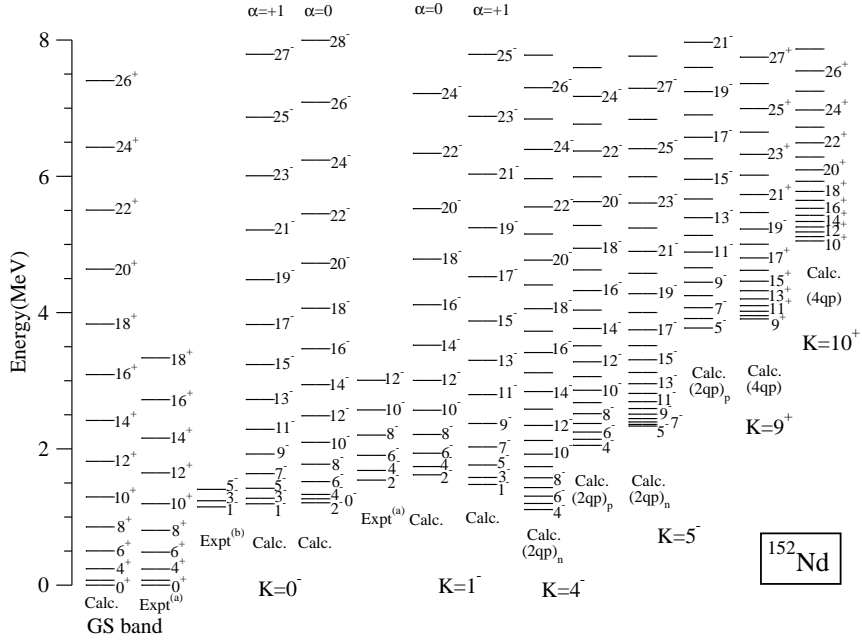

FIG. 4: Same as Fig. 2 but for $^{152}$Nd. (a) Ref. [8], (b) Ref. [40].

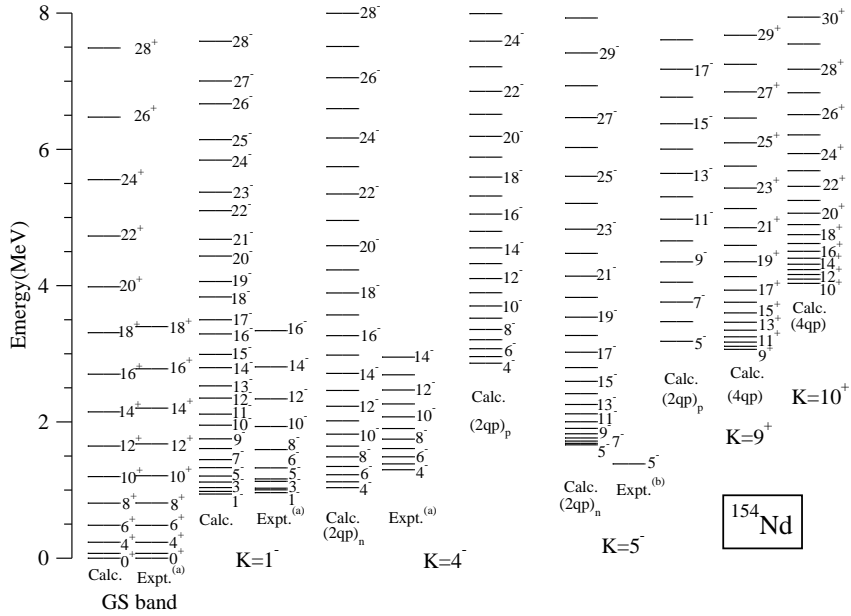

FIG. 5: Same as Fig. 2 but for $^{154}$Nd. (a) Ref. [9], (b) Ref. [10].

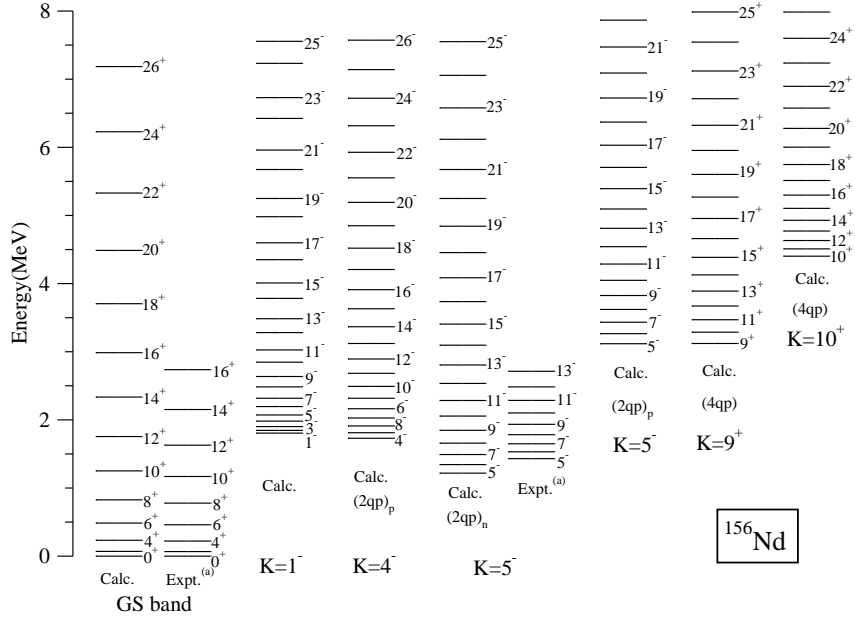

FIG. 6: Same as Fig. 2 but for $^{156}$Nd. (a) Ref. [9].

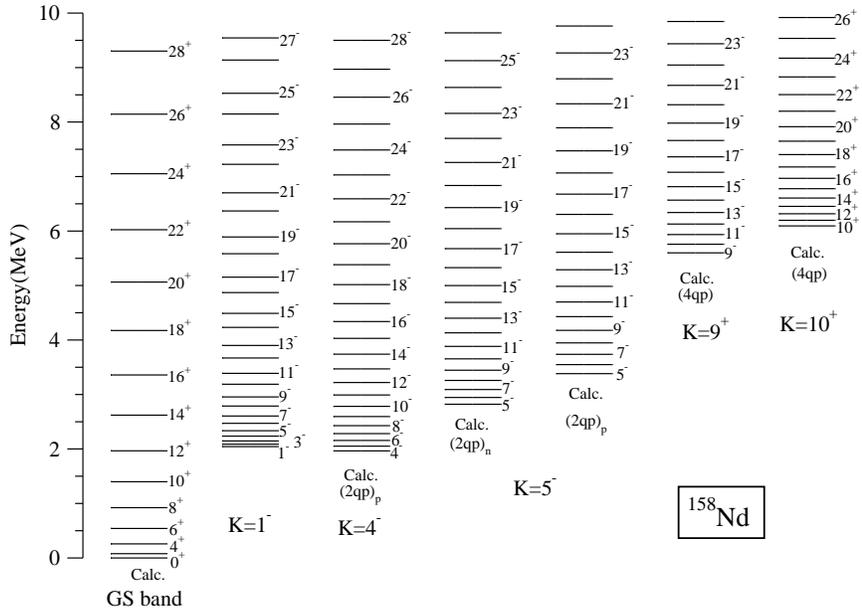

FIG. 7: Theoretical energy spectra of $^{158}$Nd.

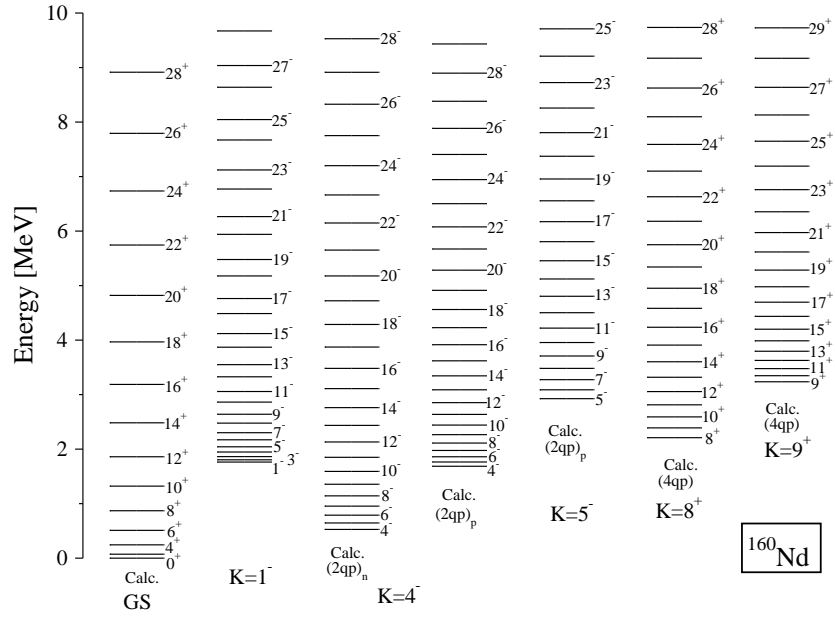

FIG. 8: Theoretical energy spectra of $^{160}$Nd.

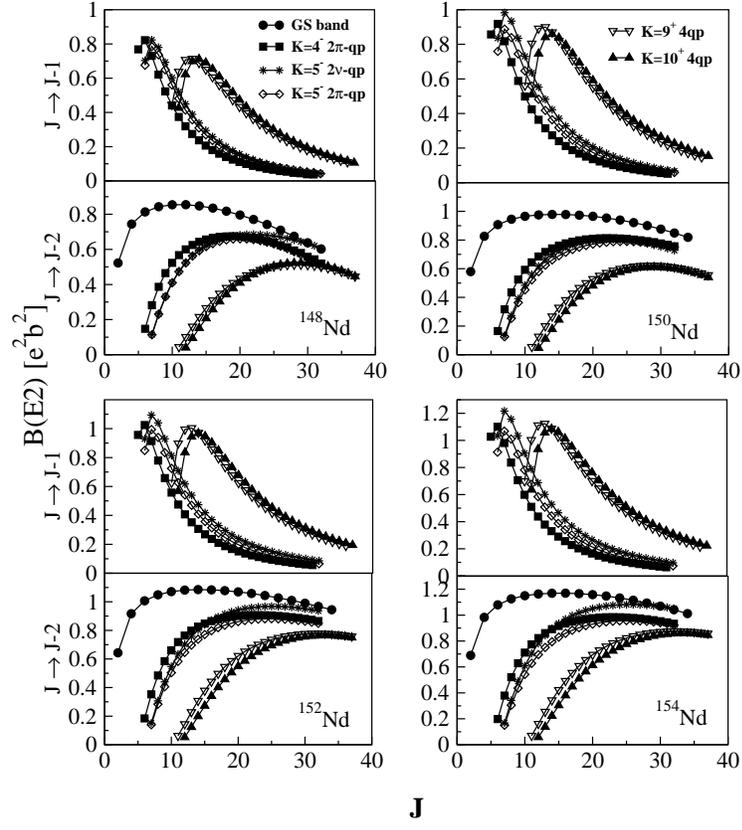

FIG. 9: B(E2) values for various bands of $_{60}$Nd nuclei.

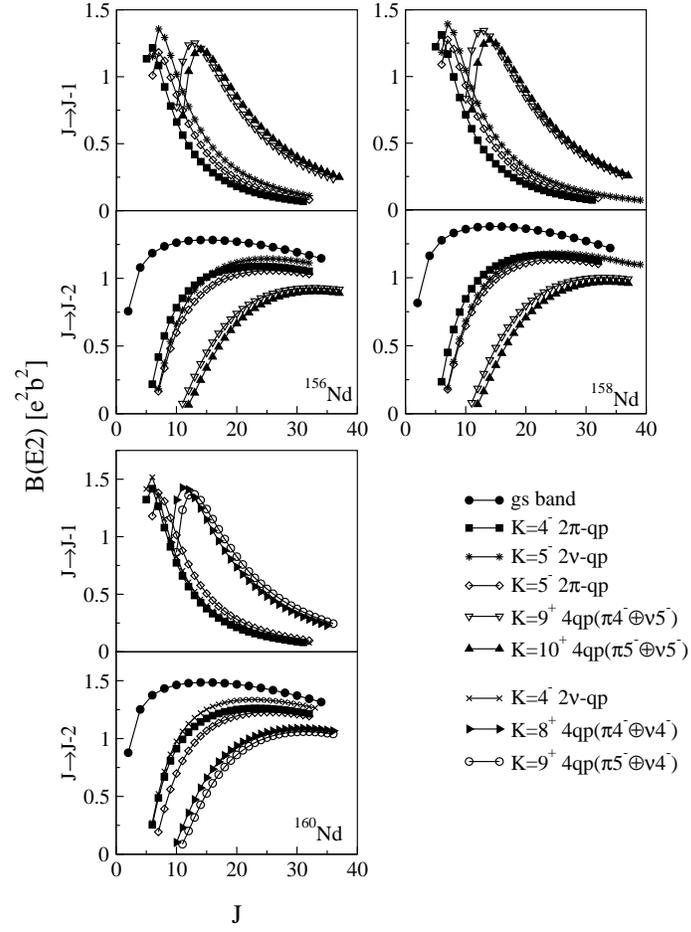

FIG. 10: B(E2) values for various bands of $_{60}$Nd nuclei.

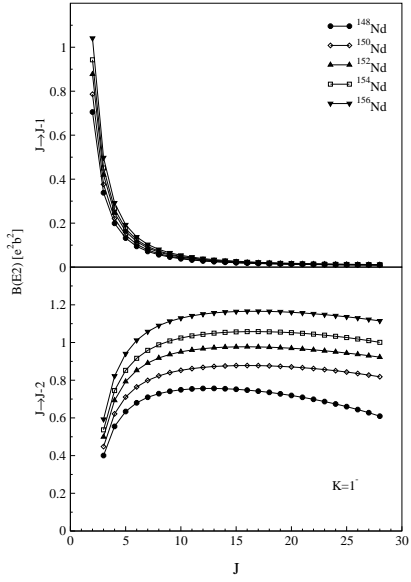

FIG. 11: B(E2) values for $K = 1^-$ band(proton excitations) of $_{60}$Nd nuclei.

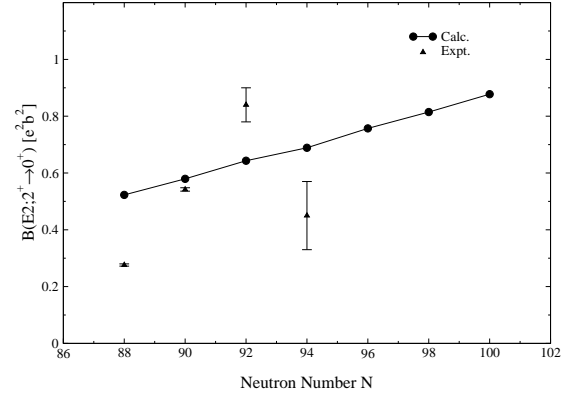

FIG. 12: Variation of B(E2;$2^+ \rightarrow 0^+$) for $_{60}$Nd nuclei. The experimental values are taken from Ref. [11].

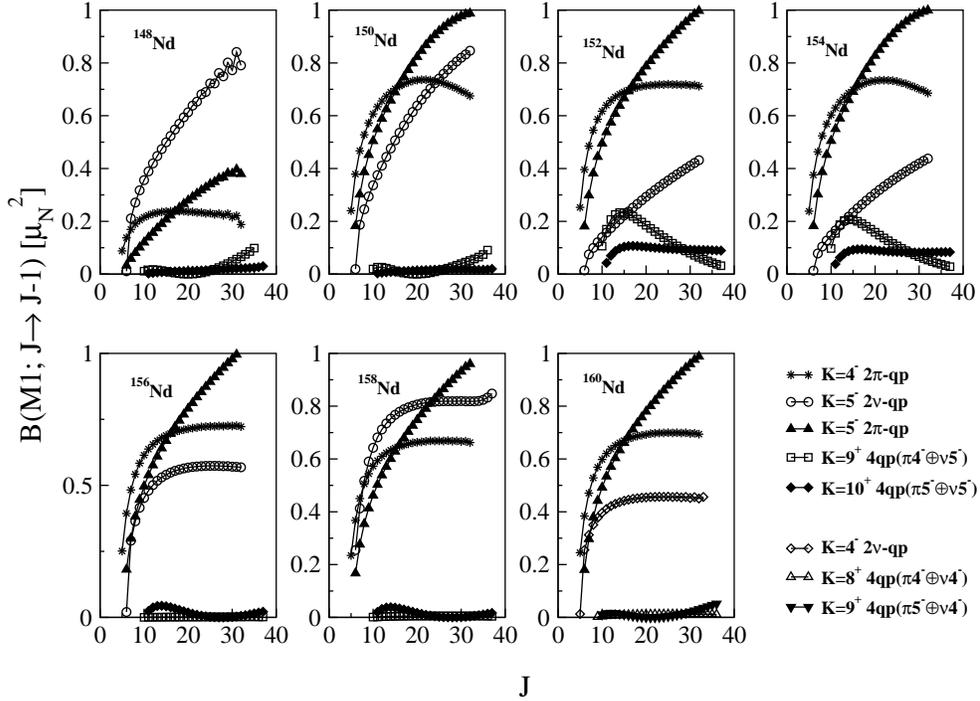

FIG. 13: B(M1) values for $_{60}$Nd nuclei.

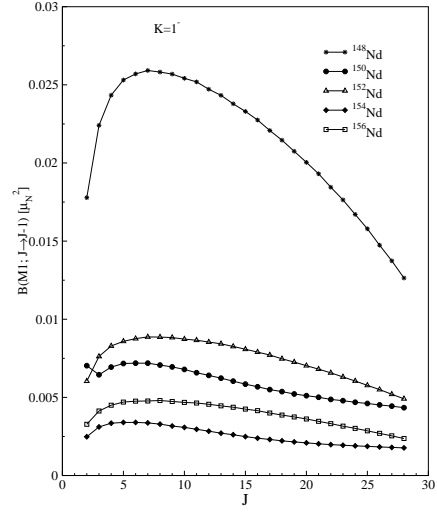

FIG. 14: B(M1) values for $K = 1^-$ band(proton excitations) of $_{60}$Nd nuclei.